\documentstyle[epsf]{l-aa}
\begin{document}

\def\clipfig#1{\def\lbracket{[}\def\testit{#1}%
    \ifx\testit\lbracket\let\next=\optclipfig\else\let\next=\stdclipfig\fi%
    \next{#1}}
%
\newcommand {\hclipfig} [7] {\clipfig[#7]{#1}{#2}{#3}{#4}{#5}{#6}}
%
\def\usemodepsfig {\global\def\cfmode{x}\typeout{*** set clipfig to PSFIG mode ***}}
\def\usemodeepsf  {\global\def\cfmode{}\typeout{*** set clipfig to EPSF mode ***}}
\def\useunitmm    {\global\def\cfunit{x}\typeout{*** set clipfig to use mm as unit ***}}
\def\useunitcm    {\global\def\cfunit{}\typeout{*** set clipfig to use cm as unit ***}}
\def\clipfigsettings {\ifx\cfmode\empty\def\ccfmode{EPSF }\else\def\ccfmode{PSFIG }\fi%
    \ifx\cfunit\empty\def\ccfunit{cm }\else\def\ccfunit{mm }\fi%
    \typeout{*** current clipfig settings: \ccfmode mode, using \ccfunit as unit ***}}
%
%
%
%
\def\stdclipfig#1#2#3#4#5#6{\ifx\cfmode\empty%
    \let\next=\eclipfig\else\let\next=\pclipfig\fi%
    \next{#1}{#2}{#3}{#4}{#5}{#6}}
\def\optclipfig#1#2]#3#4#5#6#7#8{\ifx\cfmode\empty%
    \let\next=\ehclipfig\else\let\next=\phclipfig\fi%
    \next{#3}{#4}{#5}{#6}{#7}{#8}{#2}}
%
%
%
\newcommand {\pclipfig}[6] {\ifx\cfunit\empty%
        \psfig{figure=#1.ps,width=#2cm,bbllx=#3cm,bblly=#4cm,bburx=#5cm,%
           bbury=#6cm,clip=}\else%
        \psfig{figure=#1.ps,width=#2mm,bbllx=#3mm,bblly=#4mm,bburx=#5mm,%
           bbury=#6mm,clip=}\fi}
\newcommand {\phclipfig}[7] {\ifx\cfunit\empty%
        \hspace{#7cm}\psfig{figure=#1.ps,width=#2cm,bbllx=#3cm,bblly=#4cm,%
           bburx=#5cm,bbury=#6cm,clip=}\else%
        \hspace{#7mm}\psfig{figure=#1.ps,width=#2mm,bbllx=#3mm,bblly=#4mm,%
           bburx=#5mm,bbury=#6mm,clip=}\fi}
%
%
%
\newcommand {\eclipfig}[6]{%
  \ifx\cfunit\empty\epsfxsize=#2cm\else\epsfxsize=#2mm\fi%
  \epsfclipon\epsfverbosetrue%
  \cfcmtopspts{#3}\cfllxi=\cftempi\cfllxf=\cftempf%
  \cfcmtopspts{#4}\cfllyi=\cftempi\cfllyf=\cftempf%
  \cfcmtopspts{#5}\cfurxi=\cftempi\cfurxf=\cftempf%
  \cfcmtopspts{#6}\cfuryi=\cftempi\cfuryf=\cftempf%
  \def\cfstra{\number\cfllxi.\number\cfllxf}%
  \def\cfstrb{\number\cfllyi.\number\cfllyf}%
  \def\cfstrc{\number\cfurxi.\number\cfurxf}%
  \def\cfstrd{\number\cfuryi.\number\cfuryf}%
  \hbox{\epsfbox[{\cfstra} {\cfstrb} {\cfstrc} {\cfstrd}]{#1.ps}}}
\newcommand {\ehclipfig}[7]{%
  \ifx\cfunit\empty\epsfxsize=#2cm\else\epsfxsize=#2mm\fi%
  \epsfclipon\epsfverbosetrue%
  \cfcmtopspts{#3}\cfllxi=\cftempi\cfllxf=\cftempf%
  \cfcmtopspts{#4}\cfllyi=\cftempi\cfllyf=\cftempf%
  \cfcmtopspts{#5}\cfurxi=\cftempi\cfurxf=\cftempf%
  \cfcmtopspts{#6}\cfuryi=\cftempi\cfuryf=\cftempf%
  \def\cfstra{\number\cfllxi.\number\cfllxf}%
  \def\cfstrb{\number\cfllyi.\number\cfllyf}%
  \def\cfstrc{\number\cfurxi.\number\cfurxf}%
  \def\cfstrd{\number\cfuryi.\number\cfuryf}%
  \ifx\cfunit\empty\hspace{#7cm}\else\hspace{#7mm}\fi%
  \hbox{\epsfbox[{\cfstra} {\cfstrb} {\cfstrc} {\cfstrd}]{#1.ps}}%
  \vspace{-1mm}}
%
%
%
\newdimen\cfllxi \newdimen\cfllyi  \newdimen\cfurxi  \newdimen\cfuryi
\newdimen\cfllxf \newdimen\cfllyf  \newdimen\cfurxf  \newdimen\cfuryf
\newdimen\cftemp \newdimen\cftempi \newdimen\cftempf
\newdimen\cfpspoint \cfpspoint=1bp
%
%
%
\newcommand{\cfcmtopspts}[1]{\ifx\cfunit\empty%
  \cftemp=#1cm\else\cftemp=#1mm\fi%
  \multiply\cftemp10\divide\cftemp\cfpspoint%
  \cftempf=\cftemp\divide\cftemp10\cftempi=\cftemp\multiply\cftemp10%
  \advance\cftempf-\cftemp}
%
%
\def\cfmode{}\def\cfunit{}\clipfigsettings
%

\useunitmm


\newcommand{\lb}{$\lambda$}   
\newcommand{\sm}[1]{\footnotesize {#1}}
\newcommand{\inft}{$\infty$}
\newcommand{\vlv}{$\nu L_{\rm V}$}
\newcommand{\lv}{$L_{\rm V}$}
\newcommand{\lx}{$L_{\rm x}$}
\newcommand{\lsoft}{$L_{\rm 250eV}$}
\newcommand{\lhard}{$L_{\rm 1keV}$} 
\newcommand{\vlsoft}{$\nu L_{\rm 250eV}$}
\newcommand{\vlhard}{$\nu L_{\rm 1keV}$}
\newcommand{\vlir}{$\nu L_{60\mu}$}
\newcommand{\ax}{$\alpha_{\rm x}$}
\newcommand{\aopt}{$\alpha_{\rm opt}$}
\newcommand{\aoxh}{$\alpha_{\rm oxh}$}
\newcommand{\airhard}{$\alpha_{\rm 60\mu-hard}$}
\newcommand{\aoxsoft}{$\alpha_{\rm ox-soft}$}
\newcommand{\aio}{$\alpha_{\rm io}$}
\newcommand{\aixs}{$\alpha_{\rm ixs}$}
\newcommand{\aixh}{$\alpha_{\rm ixh}$}
\newcommand{\hb}{H$\beta_{\rm b}$}
\newcommand{\nh}{$N_{\rm H}$}
\newcommand{\nhgal}{$N_{\rm H,gal}$}
\newcommand{\nhfit}{$N_{\rm H,fit}$}
\newcommand{\dnh}{$\Delta N_{\rm H}$}
\newcommand{\ale}{$\alpha_{\rm E}$}
\newcommand{\cts}{$\rm {cts\,s}^{-1}$}
\newcommand{\pl}{$\pm$}
\newcommand{\kev}{\rm keV}
\newcommand{\rb}[1]{\raisebox{1.5ex}[-1.5ex]{#1}}
\newcommand{\ten}[2]{#1\cdot 10^{#2}}
\newcommand{\msun}{$M_{\odot}$}
\newcommand{\dM}{\dot M}
\newcommand{\dMM}{$\dot{M}/M$}
\newcommand{\dMedd}{\dot M_{\rm Edd}}
\newcommand{\km }{km\,$\rm s^{-1}$}
\newcommand{\hahb}{H$\alpha$/H$\beta$}
\def\lesssim{\mathrel{\hbox{\rlap{\hbox{\lower4pt\hbox{$\sim$}}}\hbox{$<$}}}}
\def\gtrsim{\mathrel{\hbox{\rlap{\hbox{\lower4pt\hbox{$\sim$}}}\hbox{$>$}}}}

\thesaurus{03(02.16.2; 11.01.2; 11.14.1; 11.19.1)}
\title{Scattering and Absorption in Soft X-ray Selected AGN: An Optical Polarization Survey}
\author{D. Grupe\inst{1,2,}\thanks{Current address: MPE, Giessenbachstr.,
D-85748 Garching},
Beverley J. Wills\inst{1},
D. Wills\inst{1},
\and K. Beuermann\inst{2}
}
\offprints{D. Grupe (dgrupe@xray.mpe.mpg.de)}
\institute{Department of Astronomy \& McDonald Observatory, University of Texas at Austin, RLM 15.308, Austin, TX 78712, U.S.A.
\and Universit\"ats-Sternwarte, Geismarlandstr. 11, D-37083 G\"ottingen,
Germany
}
\date{Received 20.08.1997; accepted 17.10.1997}
\maketitle
\markboth{D. Grupe et al.: Optical Polarimetry of Soft X-ray AGN}{}

\begin{abstract}

We have surveyed the optical linear polarization of a completely identified sample
of 43 bright soft-X-ray-selected ROSAT AGN.  
Most (40) of these AGN show low polarization ($\lesssim$1\%), 
and no clear optical
reddening.  This supports the suggestion from rapid X-ray variability, disk-like 
spectral energy distributions, and lack of cold X-ray absorption, that we are viewing
a bare AGN disk.
IRAS\,F12397+3333 and IRAS\,13349+2438 show high polarization increasing to
the UV -- clear evidence for scattering.  As well as steep, soft-X-ray
spectra, they show optical reddening and rapid X-ray variability,
but almost no cold X-ray absorption -- a combination that suggests dusty ionized gas along
the line-of-sight.    Brandt et al. suggested and found these `warm absorbers'
for IRAS\,13349+2438.  IRAS\,F12397+3333 is a new candidate.  Combining our data
with the
optical and X-ray spectra of the high polarization narrow-line Seyfert 1 nuclei (NLSy1s) 
investigated by Goodrich reveals strong correlations among optical reddening indicators 
(\aopt\ and \hahb), [O\,III]/\hb, and cold intrinsic X-ray absorption \dnh.
Optical reddening underpredicts the cold X-ray absorption, suggesting dusty warm 
absorbers in all the highly polarized AGN.
 The existence of these scattering-polarized and reddened NLSy1s suggests an 
orientation Unified Scheme within the class of NLSy1s, analogous to that linking
Seyfert 1 and Seyfert 2 nuclei.

For some highly polarized and optically selected AGN we present new analysis of
archival X-ray data, and for the highly polarized AGN new optical spectroscopy
is presented in an appendix.

\keywords{Polarization -- Galaxies: active -- Galaxies: nuclei of
-- Galaxies: Seyfert}
\end{abstract}
\section{\label{intro}Introduction}

A popular idea among astronomers is that all AGN 
are powered by the same kind of engine: 
accretion of matter through an accretion disk onto a super-massive black hole.
This idea receives strong support from high, Keplerian velocities of gas in the
centers of active galaxies, and alignment of the axes of these gaseous disks
with powerful
radio jets.  Understanding of the basic fueling mechanism remains elusive, but
there may be new clues, related to the classic energy-budget problem and
the so-called First Principal Component, both linking the ionizing continuum
and the emission-line regions:

(a) The energy budget problem is that there appears to be insufficient ionizing 
continuum photons to power the broad emission lines, especially the 
strong, blended Fe\,II lines (e.g. Netzer 1985; Collin-Souffrin et al. 1988).  
Solutions may be that the Fe\,II arises from a different region,
excited not by photoionization but by mechanical heating of the gas 
(e.g. Norman \& 
Miley 1984; Joly 1991), or that the continuum incident on the broad 
emission-line gas is different from that observed.

(b) In AGN's UV-optical spectra there are several clear, related trends:
increasing broad-line Fe\,II emission corresponds to decreasing width (FWHM)
of
H$\beta$\ from the broad-line region (BLR), and to decreasing line emission
from
lower velocity, lower density gas -- [O\,III]\,$\lambda$5007 emission from the
narrow line region (NLR), and UV line emission from lower velocity gas of the
BLR (Boroson \& Green 1992; Laor et al. 1997; Lawrence et al. 1997;
Brotherton et al. 1997, in preparation).
ROSAT, with its high sensitivity to soft X-rays, has played an important
role in showing that these trends correspond to increasing steepness of the
soft X-ray spectra (Boller et al. 1996,  Laor et al. 1997, Grupe et al. 1997,
in preparation).
Correlation analyses show that the dominant spectrum-to-spectrum variation
in low and high redshift AGN samples can be reduced to a linear combination of
these emission line and continuum parameters -- the `First Principal Component'.
This First Principal Component points to a single, as yet mysterious, 
underlying mechanism 
relating AGN's ionizing continuum ($\sim$ 0.01-0.2 keV), the optical Fe\,II
emission, and emission from the BLR and NLR.

While rarely discovered in optical surveys, narrow-line Seyfert 1 nuclei (NLSy1s) are
found in abundance in soft-X-ray surveys.  These AGN, having broad 
H$\beta$\ emission lines with FWHM $\lesssim 2000$ km s$^{-1}$, and steep,
strong, soft-X-ray spectra, lie at the 
rich-Fe\,II extreme of the First Principal Component, and so could give us
insight into central engine mechanisms.  

One hypothesis proposed to explain their
strong, often time-variable, soft X-ray emission, their steep X-ray spectra and narrow
BLR H$\beta$\ emission lines, is that NLSy1s represent unobscured, low-mass AGN
(Boller et al. 1996; Grupe 1996; Grupe et al. 1997, 1997, in preparation).
Lower mass means lower gas
velocities, producing narrower H$\beta$\ emission from the BLR.  The strong, soft 
X-ray-emission arises from an unobscured view of the inner regions of a smaller, and 
therefore hotter, accretion disk.  This direct, rather than scattered-light, view of the
tiny emission region allows very rapid variability.
Further support for this 
hypothesis is an observed anticorrelation between optical and soft-X-ray spectral 
indices, and, in the most luminous AGN, an optical continuum slope equal to that
expected from optically thick, thermal, disk emission.  On this hypothesis, NLSy1s
should be unpolarized.

Another hypothesis suggests that NLSy1s are Seyfert 1s viewed pole-on.  Kinematic
projection of a disk-like BLR results in a narrower radial velocity distribution,
and steep X-ray spectra result from a view of the `funnel' of a geometrically thick
accretion disk (Madau 1988).

Perhaps some fraction of these NLSy1s are like the archetype of highly
polarized QSOs, IRAS\,13349+2438 (Wills et al. 1992b, Brandt et al. 1996).
The soft X-ray spectrum of this AGN is strong, variable, and steep,
with no signs of absorption by neutral gas.
In the optical, it shows typical NLSy1 characteristics -- narrow H$\beta$\ from 
the BLR, very strong Fe\,II, and very weak [O\,III]\,$\lambda$5007.  In this AGN there
are two views to the central continuum source and BLR -- one a direct view that is 
partially obscured by dusty hot gas in the torus, and the other a less obscured, but
indirect, view in which the spectrum is scattering-polarized by electrons or dust in
the opening cone of the torus.  The unusual combination of UV-optical
reddening and lack of cold X-ray absorption was presented by Brandt et al. (1996) as
evidence for dusty, ionized gas near the nucleus.  Discovery of more such objects would
allow further optical-UV-X-ray investigation of this newly discovered constituent of
AGN gas.

The above pictures for NLSy1s are not mutually exclusive.  In axisymmetric
Unified Schemes, a low-mass or thick-disk central engine may be surrounded by a 
dusty nuclear torus.
A low-inclination polar view may reveal an unobscured central engine,
while a higher inclination line-of-sight may graze the dusty nuclear torus, passing
through ionized gas (e.g. IRAS\,13349+2438).
As dust further dims the direct view of the nucleus, the scattered,
polarized AGN spectrum becomes more prominent.  At even higher inclinations, cold
dusty gas blocks the direct view completely, but a few percent of the NLSy1 spectrum
could be scattered towards the observer.  Even in a high inclination view, low
density NLR gas, ionized
on kpc scales, may reveal the presence of a buried NLSy1 in total light.
The well-known ``Seyfert 1 nucleus buried within the prototype Seyfert 2'' galaxy
NGC\,1068, the Rosetta Stone of Unified Schemes (Antonucci \& Miller 1985),
may be an excellent high-inclination example with the parsec-scale AGN seen only in
scattered light -- a faint, polarized spectrum with UV-optical continuum, 
strong Fe\,II blends, and BLR H$\beta$\ intrinsic
FWHM $\sim$2900 \km (Miller et al. 1991), together with a very steep, unabsorbed soft 
X-ray spectrum (\ax\ $\sim$2.4), showing no variability on several-year time-scales
(Elvis \& Lawrence 1988; Marshall et al. 1993; Smith et al. 1993).

In this paper we have used broad-band polarimetry and spectroscopy of a ROSAT 
sample of soft-X-ray-selected AGN to relate scattering and absorption in the
optical and X-rays, in order to investigate anisotropic emission, kinematics and
dusty gas distribution in these special AGN.
In Sect. \ref{obs} we describe the sample, as well as observations and data 
reduction for the optical polarimetry, ROSAT X-ray spectroscopy, and optical
spectroscopy.
Sect. \ref{res} presents the results, including some correlations.  Here we
provide more detailed information for the highly polarized sources, and collect together
comparable X-ray and optical data for the AGN of Goodrich's (1989) 
spectropolarimetry study of optically selected NLSy1s.
Sect. \ref{discus} discusses the absorption and scattering properties of the
sample as a whole, and for individual sources, including a comparison with the
optically selected NLSy1s.  A summary is given in Sect. \ref{sum}.  Appendix
A shows our spectra for all the highly polarized sources discussed in this paper.

\section{\label{obs} Observations and Data}

\subsection{\label{obs,samp}The Soft X-ray Sample}

The sample of bright, soft-X-ray AGN (Grupe et al. 1997) was selected
from the ROSAT All-Sky Survey (RASS, Voges et al. 1993, 1997).  This deep
survey, using the Position Sensitive Proportional Counter (PSPC, Pfeffermann
et al. 1986), was the first to extend to soft-X-ray photon energies $\sim$\
0.1 keV.  Our X-ray sample is completely identified, comprising all 95 AGN with
RASS PSPC count rates $\gtrsim$0.5 $\rm cts~s^{-1}$, hardness ratio HR1 
$\lesssim$0.0, and $|b|>20^{\circ}$.  HR1 is defined as the ratio of the 
difference to the sum of hard and soft counts of the source, where the hard and
soft bands are defined between 0.4-2.4 keV and 0.1-0.4 keV.  This HR1 limit
corresponds to a steep soft X-ray spectrum with \ax $\gtrsim 1.5$\ for 
\nh$\sim 2 \times 10^{20}$\,cm$^{-2}$.
The bright, soft X-ray selection results in a sample with very little or no
neutral hydrogen absorption.  The X-ray and optical properties of the sample
were presented and discussed by Grupe (1996) and Grupe et al. (1997, and 1997 
in preparation).
About half the sample are NLSy1s.  The median FWHM for broad H$\beta$\ is
2250 \km, with a 90-percentile range from $\sim$1300 \km~ to 4200 \km.
We restrict the present polarization survey and analysis to the 43 northern
sources ($\delta > 0^{\circ}$).

\subsection{\label{obs,pol}Polarimetry Observations}

For the soft X-ray sample, linear polarization was measured using a 
broad-band polarimeter (Breger 
1979) on the Struve 2.1-m telescope at McDonald Observatory.  The detector
was a cooled ($\sim -20$C) Hamamatsu R943-02 phototube with extended red
sensitivity [GaAs(Cs) photocathode] operated in photon-counting mode.
A new interface and data acquisition computer resulted in improved noise
characteristics with uncertainties consistent with photon statistics.
Otherwise, the filters,
observational technique (a few 200-second integrations on-source,
interleaved with background integrations), calibrations, and reductions
were as described by Wills et al. (1992a).  All measurements were made with
an aperture of projected diameter 7.4\arcsec.
We were interested in searching for scattering polarization, so
we aimed for 1$\sigma$\ measurement uncertainties $\lesssim$0.4\%, because
interstellar polarization mechanisms may become important at that level,
and are then difficult to distinguish from scattering, based on degree of
polarization and wavelength-dependence.
In some cases where the measurements indicated real polarization we 
estimated the contribution from Galactic interstellar polarization
by measuring the polarization of stars nearby in projection on the sky.
In order to probe as long a path length as possible through
our galaxy we chose (where possible) bright, blue stars, preferably known
to be of early spectral type.
Of the 43 northern AGN, 26 were observed by us and data for 17 were
taken from the literature.

\subsection{\label{obs,x}X-ray data}

The RASS data have been used for the statistical investigation of the 
soft-X-ray sample, except for well-known AGN.  Details of the reductions 
are given by Grupe (1996) and Grupe et al. (1997).  For the 19 well-known
sources we retrieved pointed observation data from the ROSAT public archive
at MPE Garching.  The January 1996 version of the EXSAS package (Zimmermann
et al. 1994) was used for data reduction and analysis.  The X-ray spectral
index \ax~ ($F_{\nu} \propto \nu^{-{\alpha_x}}$, the photon index 
$\Gamma=1+$\ax) was determined in the
energy range 0.1 -- 2.4 keV using single power-law fits, assuming neutral
absorption given by Milky Way HI 21cm column densities $N_{\rm H,gal}$\
(Dickey \& Lockman 1990) with abundances given by Morrison \& McCammon (1983).
Uncertainties in
\ax\ are typically 0.1 to 0.5, depending on X-ray brightness.  Grupe et al.
(1997) also made fits to the X-ray spectra with unconstrained neutral 
hydrogen column-density $N_{\rm H}$.  Although the latter was rarely 
significantly different from Galactic values, we measured the difference
\dnh $= N_{\rm H} - N_{\rm H,gal}$\ as a potential indicator of intrinsic
neutral hydrogen absorption.
For those objects for which no X-ray spectra were readily available, we 
used the hardness ratios HR1 and HR2 given in the bright source catalogue
(Voges et al. 1997).  HR2 is like HR1, but with hard and soft bands defined
between 0.9--2.0 keV and 0.5--0.9 keV.  
The method is to simulate count-rate spectra for a grid of \ax\ and \nh~values, 
using the known instrumental response.  From these spectra we calculate
HR1 and HR2 for each combination of \ax\ and \nh.  Thus we derive
the unique combination \ax\ and \nh\ that reproduce the observed HR1 and HR2.
Uncertainties in \dnh\ are similar to those in $N_{\rm H}$\ determined
from the fits with $N_{\rm H}$\ unconstrained.  The standard deviations for 
individual values
are about 0.8 $\times 10^{20}$\ cm$^{-2}$\ and 1.2 
$\times 10^{20}$\ cm$^{-2}$\,
as determined from the median and mean of the distribution of standard
deviations given from the fitting.

For the highly polarized AGN for which ROSAT pointed observations were available, the
data were analyzed in the same way as for the RASS data.

\subsection{\label{obs,opt}Optical Spectroscopy} 

In general, we used the medium-resolution optical spectra (FWHM $\sim$5\AA)
from Grupe (1996) and Grupe et al. (1997, in preparation), obtained in March 1994.
The highly polarized objects were re-observed, including H$\beta$\ and H$\alpha$, in
March 1997 with the 2.1m telescope at McDonald Observatory, but with resolution
FWHM $\sim$ 7\AA.  Spectra for the PG quasars (resolution 3.6\AA\ FWHM) are from 
observations with the 2.7-m telescope at McDonald Observatory in February
1996.  A spectrum for RE\,J1034+39 is from Puchnarewicz et al. (1995a), and
for CSO 150, from Bade et al. (1995).
Except for QSO\,1136+579, and the higher redshift AGN RX\,J1014.0+4619 and 
RX\,J1050.9+5527, we were able to find spectra in the literature:
MS\,0919.3+5133 (Stephens 1989), EXO\,1128.1+6908 (Bedford et al. 
1988), UM\,472 (Salzer et al. 1989), and NGC\,4593 (Dietrich et al. 1994).

The emission line measurements and their uncertainties are presented and 
described in detail
by Grupe et al. 1997, in preparation).  Some representative $1 \sigma$\ errors are given
in later tables.
The blended optical Fe\,II emission was measured and subtracted using a
I\,Zw\,1 template like that used by Boroson \& Green (1992).
All other measurements were made from these Fe\,II-subtracted spectra.
To determine the rest-frame equivalent width of Fe\,II 
we measured the flux
in the template between rest wavelengths 4250\AA~ and 5880\AA, after scaling
it to match the object spectra, and then divided by
the continuum flux density at 5050\AA.  In the following, H$\beta$\ refers to
the whole H$\beta$\ line.  \hb\ refers to
the broad component of H$\beta$, obtained after subtraction of the narrow
component by using the [O\,III]$\lambda5007$\ velocity profile as a template.
This method subtracts as much of the template as possible while still 
retaining a sensible shape to the peak of \hb.  For sources where the 
narrow-line emission dominates we found an [O\,III]/H$\beta_{\rm narrow}$\ 
intensity ratio
of about 10, which is not significantly different from the observed 
median in Seyfert 1s and Seyfert 1.5s (Koski 1978; Cohen 1983), providing
support for our method.

To determine the H$\alpha$/H$\beta$ intensity ratio, we used the total flux in the lines,
because the broad and narrow components are difficult to separate for the 
H$\alpha$\ blend.  In general, to subtract the fluxes of the 
[N\,II]\,$\lambda\lambda$6548,6584 lines from the H$\alpha$\ blend, it was quite
adequate to subtract 35\% of the [O\,III]$\lambda$5007 line flux from it
(Ferland \& Osterbrock 1986).  For the highly polarized sources, as we see later,
reddening of
the NLR emission could be important.  So for these we used velocity templates
derived from [O\,III] and total H$\beta$\ lines to disentangle the 
H$\alpha$--[N\,II] blend.

Optical continuum spectral indices \aopt\ were measured between 
rest-frame wavelengths of 4400\AA\ and 7000\AA.

\subsection{\label{obs,good}The Optically Selected Sample}

For comparison with our soft-X-ray-selected sample, we have compiled
polarization, spectroscopic and X-ray data for the optically selected
sample of NLSy1s that Goodrich (1989) investigated by
spectropolarimetry.  While the selection of that sample is based on
a mixture of spectroscopic criteria
(narrow H$\beta$\ emission, strong Fe\,II emission, 
[O\,III]$\lambda$5007/H$\beta$\ $< 3$, and presence of high-ionization
lines) it is unbiased with respect to soft-X-ray properties 
(\ax\ or \dnh).  Measurements were
made, as far as possible, in the same way as for our sample.
Emission-line measurements were derived from literature data, except for
the more highly polarized sources for which we obtained new 
spectroscopic data including both the H$\beta$\ and H$\alpha$\ regions
(Appendix A).  

\section{\label{res} Results}
\subsection{\label{res,x}The ROSAT Soft X-ray AGN Sample}

\begin{figure}
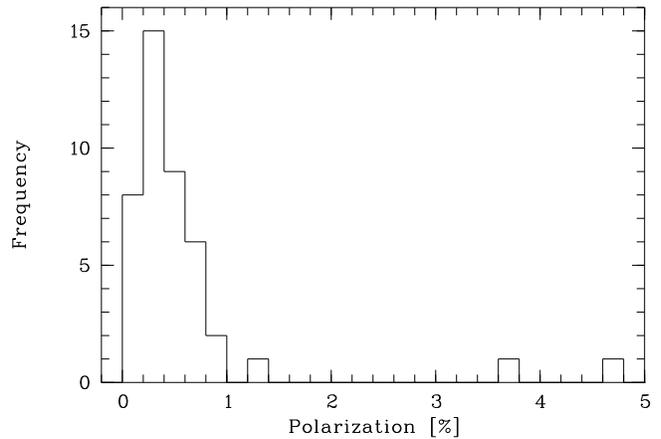

\parbox[h]{8.7cm}{
\clipfig{distr_pol}{87}{15}{15}{280}{195}}
\caption[ ]{\label{distr_pol}
Distribution of the polarization of the 43 northern soft-X-ray AGN.
}
\end{figure}

\begin{figure}
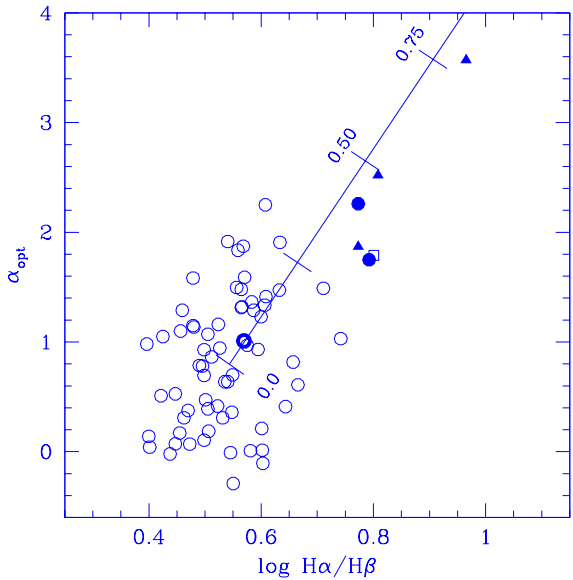

\clipfig{hahb_aopt}{87}{05}{60}{205}{240}
\caption{\label{hahb_aopt}
The optical spectral index \aopt~ vs. total \hahb.  Data for our northern and southern
X-ray-selected samples are shown as circles, with IC\,3599 shown as an open
square.  See Table \ref{res_highpol} for representative $1 \sigma$\ uncertainties.
The majority, thought to be unreddened, are shown as open circles.
The highly polarized AGN (p\,$>$1\%) are shown as filled symbols -- circles
for the soft-X-ray AGN, triangles for the optically-selected sample.  CBS\,126,
with p$\sim$1\%, is shown as a heavy circle. 
The tick-marked line in the plot displays
values of $E_{\rm B-V}$ for equal line and continuum reddening.
Polarization data are not available for the majority of southern sources.
}
\end{figure}

\begin{figure}
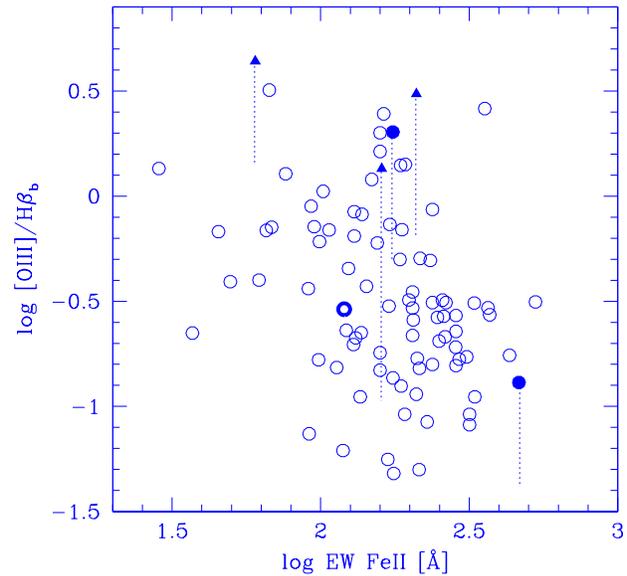

\clipfig{ewfe2_o3}{87}{05}{60}{205}{240}
\caption{\label{ewfe2_o3}
Line ratio [O\,III]/\hb~ vs. equivalent width of the optical Fe\,II blend.  
See Table \ref{res_highpol} for representative $1 \sigma$\ uncertainties.
The symbols are as defined in Fig. \ref{hahb_aopt}.  The dotted vertical lines
show the de-reddening paths, based on E(B-V) values derived from 
Fig. \ref{hahb_aopt}.
}
\end{figure}

The polarimetry data are given in Table \ref{res_pol}, where
we list the object's RX\,J name derived from the RASS J2000 coordinates,
other name, the filter used,
degree of polarization p\%, and the polarization position angle (E vector, 
measured on the sky plane in degrees from north through east), and a literature
reference where relevant.  Errors in p\% are 1$\sigma$, based on 
photon-counting statistics.  The corresponding position-angle error may be
estimated from 28.65$^{\circ} \times$\ the fractional error in p\%.
No corrections have been applied for positive 
bias in p\%.  Neither have we made any attempt to correct for unpolarized
starlight from the host galaxy.
Table \ref{star_pol} gives measurements we made for 
interstellar polarization.  In no case was this large enough to justify a
correction to our AGN polarizations.

\begin{table*}
\caption{\label{res_pol} Polarimetry of Soft X-ray-selected AGN.} 
\begin{tabular}{clclrll}
\hline
\noalign{\smallskip}
Object & Other & & \multicolumn{2}{c}{Polarization} &  
\\
RX J & Name & \rb{Filter} & 
p\% & Angle&($^{\circ}$)  &  \rb{Source of Polarimetry data} \\
\noalign{\smallskip}\hline\noalign{\smallskip}
0922.8+5121 & MS\,0919.3+5133 & none & 0.75\pl0.46 & 27 &&  McD2.1 March 1997   \\
0925.2+5217 & Mkn\,110 & none & 0.21\pl0.13  & --- && Berriman et al. (1990) \\
0956.8+4115 & PG\,0953+414 & none &  0.25\pl0.22 & --- &&  Berriman et al. (1990) \\
1004.0+2855 & PG\,1001+292 &  none & 0.77\pl0.22  & --- && Berriman et al. (1990) \\  
1005.7+4332 &            & none & 0.40\pl0.23  & 131 
&& McD2.1 April 1996 \\
1013.0+3551 & CBS\,126 & none & 1.26\pl0.13  & 112 
&& McD2.1 April 1996  \\
& & CuSO4 & 1.19\pl0.12  & 118  && McD2.1 April 1996  \\
1014.0+4619 &            & none & 0.03\pl0.31  & 177 
&&  McD2.1 April 1996 \\
1017.3+2914 &            & none & 0.42\pl0.16  & 161 
&&  McD2.1 April 1996 \\
1019.2+6358 & Mkn\,141 & 3800 - 5600 \AA   & 0.08\pl0.28  & --- && Berriman (1989) \\
& & 3800 - 5600 \AA &  0.29\pl0.28 & ---   && Martin et al. (1983) \\
1025.5+5140 & Mkn\,142 &  none  & 0.11\pl0.27  & --- && Berriman et al. (1990) \\
1034.6+3938 & RE\,J1034+39 & none  & 0.15\pl0.21 & 17 && McD2.1 March 1997 \\
& & 3900 - 5600 \AA & 0.36\pl0.13 & 78 & & Breeveld \& Puchnarewicz (1997)\\
& & 5800 - 9200 \AA & 0.40\pl0.13 & 71 & & Breeveld \& Puchnarewicz (1997)\\
1050.9+5527 &            & none & 0.31\pl0.27  & 64 &&  McD2.1 April 1996 \\
1058.5+6016 & EXO\,1055+60 & none & 0.44\pl0.26  & 126 && McD2.1 April 1996 \\
1107.2+1628 & PKS\,1104+16 & none  & 0.63\pl0.14 & 157 && Wills et al. (1992a) \\
1117.1+6522 &            & none & 0.40\pl0.22 & 17 && McD2.1 April 1996 \\
1118.5+4025 & PG\,1115+407 & none & 0.37\pl0.24  & --- && Berriman et al. (1990) \\
1119.1+2119 & Ton\,1388 & none  & 0.23\pl0.11 & --- && Berriman et al. (1990) \\ 
1121.7+1144 & Mkn\,734 & 3800 - 5600 \AA  & 0.32\pl0.13  & --- && 
Berriman (1989), Martin et al. (1983) \\
& & none &  0.39\pl0.17  & --- &&  Berriman et al. (1990) \\
1131.1+6851 & EXO\,1128.1+6908 & none & 0.29\pl0.18  & 37 && McD2.1 March 1997 \\
1138.8+5742 & QSO\,1136+579 & none  & 0.33\pl0.28 & && McD2.1 March 1997 \\ 
1139.2+3355 & Z 1139+34 & none & 0.48\pl0.16 & 149 && McD2.1 April 1996 \\
1145.1+3047 & CSO\,109 & none & 0.03\pl0.19  & 172 && McD2.1 April 1996 \\
1203.1+4432 & NGC\,4051 &  3800 - 5600 \AA  & 0.66\pl0.10  & && 
Berriman (1989), Martin et al. (1983) \\
1203.5+0229 & UM\,472 &  none & 0.62\pl0.42  & 41 && McD2.1 March 1997 \\
1214.3+1403 & PG\,1211+143 & none &   
0.07\pl0.09 &  --- && Berriman et al. (1990) \\
1231.6+7044 &            &  none & 0.86\pl0.23 & 170 && McD2.1 April 1996 \\
1233.6+3101 & CSO\,150 &  none & 0.20\pl0.26  & 175 && McD2.1 March 1997 \\
1237.7+2642 & IC\,3599 & none & 0.24\pl0.33 & 17 && McD2.1 April 1996 \\
1239.6$-$0520 & NGC\,4593 & 3800 - 5600 \AA   & 0.49\pl0.16  & --- && 
Berriman (1989), Martin et al. (1983) \\
1242.1+3317 & IRAS\,F12397+3333 & none & 3.77\pl0.20 & 79.6 && McD2.1 April 1996 \\
& &  U & 7.00\pl0.57  & 82.1 && McD2.1 April 1996 \\
& &  CuSO4 & 5.20\pl0.27  & 80 && McD2.1 April 1996 \\
& &  B & 4.99\pl0.38  & 81 && McD2.1 April 1996 \\
& &  V & 3.98\pl0.30  & 78.7 && McD2.1 April 1996 \\
& &  R & 3.50\pl0.23  & 80.9 && McD2.1 April 1996 \\
& &  I & 2.16\pl0.33  & 84.4 && McD2.1 April 1996 \\
1312.9+2628 &            & none & 0.20\pl0.29  & 103 && McD2.1 April 1996 \\
1314.3+3429 &            & none & 0.30\pl0.15  & 88 && McD2.1 February 1995
\\
1323.8+6541 & PG\,1322+659 & none  &  
0.81\pl0.22  & --- && Berriman et al. (1990) \\
1337.3+2423 & IRAS\,13349+2438  & none & 
4.63\pl0.08  & 124.5 && Wills et al. (1992b) \\ 
1355.2+5612 &            & none & 0.53\pl0.22  & 86 && McD2.1 April 1996 \\
1405.2+2555 & PG\,1402+261 & none  & 0.29\pl0.14 &  && Berriman et al. (1990) \\
1413.6+7029 &            & none & 0.20\pl0.45  & 133 && McD2.1 April 1996 \\
1417.9+2508 & NGC\,5548 & 3800 - 5600 \AA  & 0.71\pl0.10  & --- && 
Berriman (1989), Martin et al. (1983) \\
1431.0+2817 & Mkn\,684 & none  & 0.18\pl0.04  & 86 && Goodrich (1989) \\
1442.1+3526 & Mkn\,478 & 3800 - 5600 \AA  & 0.43\pl0.15  & --- && Berriman (1989) \\
& & 3800 - 5600 \AA & 0.46\pl0.15  & --- && Martin et al. (1983) \\
& & none  & 0.26\pl0.17  &  --- && Berriman et al. (1990) \\
1618.1+3619 &            & none & 0.35\pl0.35  & 22 && McD2.1 February 1995
\\
1627.9+5522 & PG\,1626+554 & none & 0.59\pl0.19  & --- && Berriman et al. (1990) \\
1646.4+3929 &            & none  & 0.39\pl0.26  & 0 && McD2.1 April 1996 \\
\noalign{\smallskip}\hline\noalign{\smallskip} \\
\end{tabular}
\end{table*}

\begin{table*}
\caption{\label{star_pol} Measurements for interstellar polarization. 
Right ascension and declination offsets $\Delta \alpha$\ and $\Delta \delta$\
in arcminutes relative to the AGN coordinates
 }
\begin{tabular}{llrrccr}
\hline
\noalign{\smallskip}
AGN & Star & & & & \multicolumn{2}{c}{Polarization} 
\\
Name & Name & \rb{$\Delta \alpha$} & \rb{$\Delta \delta$} 
&  \rb{Filter} & \%  & Angle \\
\noalign{\smallskip}\hline\noalign{\smallskip}
CBS 126 & SAO 61959 & +37.4 & $-$41.8 & none & 0.10\pl0.03 & 177 \\
& Star nf & +0.8 & +0.3 & none & 0.10\pl0.09 & 7 \\ \noalign{\smallskip}
RX J1231+70 & SAO 07580 & +1.3 & +86.5 & none & 0.11\pl0.05 & 55 \\
\noalign{\smallskip}
IRAS\,F12397+3333 & SAO 63174 & +33.0 & +141.6 & none & 0.04\pl0.06 &
12 \\
& SAO 63318 & +271.2 & $-$136.1 & V & 0.03\pl0.03 & 126 \\
& & & & R & 0.06\pl0.02 & 153 \\
& & & & I & 0.09\pl0.02 & 20 \\
\noalign{\smallskip}\hline\noalign{\smallskip} \\
\end{tabular}
\end{table*}

Figure \ref{distr_pol} displays the distribution of the 
polarization.  Most of the soft-X-ray AGN are of low polarization
($\lesssim$1\%).  Three of the 43 objects show p $>$1\%.
While we have chosen 1\% as a practical cut-off for investigating scattered-light
polarization, we note that two other AGN, with detailed results from the literature,
show low but significant polarization, above interstellar levels.  One is
the well-known NLSy1 NGC\,4051.  The other is the broad-lined NGC\,5548, with 
HR1 $\sim$0.

Because one motivation of this study was to investigate Unified Schemes for NLSy1s,
we ask whether there are systematic differences between high and low polarization
AGN.  The FWHM(\hb) of the highly polarized AGN are not extremely narrow
but typical for the soft X-ray sample.
Also, \ax\ $\sim$1.7 for the highly polarized AGN, so their soft X-ray spectra tend to
be flat compared with most of our sample; but steep compared with ROSAT spectra
of optically-selected Seyfert 1 nuclei, for which \ax $\sim 1.0$\ -- 1.7
(Grupe et al. 1997).

Reddening of the broad lines and continuum is another test for whether we have
a direct view to a bare nucleus, with the narrow \hb~ and steep \ax~ being
intrinsic properties, or whether soft-X-ray-selected AGN with narrow \hb~ may
be viewed through dusty gas near the nucleus, as
in orientation Unified Schemes.  In Fig. \ref{hahb_aopt} we compare \aopt\ and
the Balmer decrement, as indicators of reddening.  The low polarization AGN
(most of the open circles) have small
\hahb\ ratios, and flat optical spectra (small \aopt), and show no relation
between \hahb\ and \aopt\ outside the expected relation resulting from
systematic calibration uncertainties, so reddening
is unlikely to be important for them.  While the low polarization sources
do not show significant reddening, the highly polarized sources do.  All the
polarized AGN (filled symbols)
have Balmer decrements $\gtrsim$6 (except CBS\,126, which we discuss later).
The one low polarization source with Balmer decrement $>$5.8 is IC\,3599, which is
thought to be a Seyfert 2, or more likely, a starburst nucleus (Sect. 
\ref{discus,ic}) and therefore different from the rest
of our sample.
Another test is to look for intrinsic absorption \dnh\ in the 
X-ray spectra.  As remarked before, RASS data did not show significant
absorption in soft X-rays for any AGN, except IC\,3599.  However,
for some of the highly polarized AGN
the higher signal-to-noise pointed observations do show probable absorption
(Sect. {\ref{res,pol})

We can also test how the highly polarized sources might be related to the First
Principal Component.  In Fig. \ref{ewfe2_o3} we show the line ratio 
[O\,III]/\hb~ vs. equivalent width of the optical Fe\,II blend, EW(Fe\,II).
The low polarization sources of the present sample show the well known inverse
[O\,III] -- Fe\,II trend seen for radio-quiet AGN as a whole
(Boroson \& Green 1992, Grupe et al. 1997, in preparation).
Our high polarization sources appear to lie at larger [O\,III]/\hb\ or larger
EW(Fe\,II) than the low polarization sources (see Sect. \ref{discus}).




\subsection{\label{res,good}Comparison with an Optically Selected NLSy1 Sample}

\begin{table*}
\caption{\label{res_goodrich} Properties of Goodrich's optically
selected NLSy1 sample}
\begin{tabular}{lccrccccrc}
\hline
\noalign{\smallskip}
Common & &  \multicolumn{2}{c}{Polarization} & CR$^i$ & & & \dnh$^i$ 
& FWHM(H$\beta_{\rm b}$)$^{ii}$ \\
Name & \rb{RX J$^i$} &   
\% & Angle & $\rm cts~s^{-1}$ & \rb{HR1$^i$} &
\rb{\ax$^i$} & $\rm 10^{20}~cm^{-2}$ & $\rm km~s^{-1}$ &
\rb{$\rm \frac{[O\,III]}{H\beta_{\rm b}}$$^{^{^{ii}}}$} \\
\noalign{\smallskip}\hline\noalign{\smallskip}
Mkn\,957 & 0041.9+4021 & 0.62\pl0.06 & 43 & 0.09 & +0.62 & 1.5 (a) 
& +0.0 (a) & 685 (a) & 0.40 (a)   \\
Mkn\,359  & 0127.5+1910 & 0.46\pl0.02 & 112 & 0.61 & +0.49 & 
1.4 (b) & +0.3 (b) & 1350 (b) & 1.73 (a)  \\
Mkn\,1044  & 0230.0$-$0859 & 0.52\pl0.05 & 144 & 2.14 & $-$0.06 & 2.0 (b)
& 
 +1.1 (b) & 1500 (c) & 0.22 (a)  \\
Mkn\,1239  & 0952.3$-$0136 & 3.35\pl0.02 & 130 & 0.05 & +0.81 & 2.9 (b)  
& +4.4 (b) & 1050 (d) & 2.57 (d)   \\
PG\,1016+336 & 1019.1+3322 & 0.20\pl0.09 &  75 & ---  & --- & --- & ---
& 1310 (a) & 0.10 (a)  \\
Mkn\,42  & 1153.7+4612 & 0.37\pl0.11 & 36 & 0.19 & $-$0.15 & 
1.6 (b) & +0.6 (b) & 670 (e) & 0.35 (a)   \\
Mkn\,766 & 1218.4+2948 & 2.34\pl0.02 & 90 & 4.71 & $-$0.02 & 1.5 (b)  
& +0.4 (b) & 1360 (d) & 3.82 (d)   \\
NGC\,4748 & 1252.2$-$1324 & 0.12\pl0.04 & 83 & 0.97 & +0.20 & 1.5 (a) 
& +0.0 (a) & 1470 (a) & 1.60 (a) \\
Mkn\,783  & 1302.9+1624 & 0.25\pl0.07 & 2 & 0.29 & +0.88 & 1.3 (a)  
& +2.8 (a) & 1900 (e) & 9.42 (a)   \\
Mkn\,684 & 1431.0+2817 & 0.18\pl0.04 & 86 & 0.58 & $-$0.23 & 1.5 (c) 
& +0.4 (c) & 1690 (f) & 0.16 (f)  \\
IRAS\,15091$-$2107  & 1511.9$-$2119 & 4.61\pl0.03 & 62 & 0.37 & +0.96 & 1.6
(d)  
& +26.1 (d) & 2250 (d) & 1.16 (d)  \\
Mkn\,291   & 1555.1+1911 & 0.23\pl0.06 & 131 & --- & --- & 
1.1 (b) & $-$1.4 (b) & 700 (g) & 0.98 (a)     \\
Mkn\,493  & 1559.1+3501 & 0.26\pl0.07 & 87 & 0.52 & $-$0.24 & 
1.7 (b) & +0.6 (b) & 1360 (d) & 0.35 (d)   \\
VII\,Zw\,742 & 1747.0+6836 & 0.20\pl0.10 & 84 & 0.21 & +0.25 & 1.3 (a)   
& $-$1.2 (a) & 1260 (a) & 0.53 (a) \\
Mkn\,507 & 1748.6+6842 & 0.61\pl0.03 & 12 & --- & --- & 0.6 (b)  
& +0.5 (b) & 965 (a) & 0.73 (a)   \\
Akn\,564  & 2242.6+2943 & 0.40\pl0.02 & 90 & 0.38 & +0.42 & 2.4 (b)
& +1.2 (b) & 1000 (b) & 1.10 (a)  \\
Mkn\,1126 & 2300.8$-$1255 & 0.47\pl0.04 &173 & 0.35 & +0.14 & 
1.5 (a) & $-$0.2 (a) & 2500 (e) & 4.15 (a)  \\
\noalign{\smallskip}\hline\noalign{\smallskip} \\
\end{tabular}
\begin{minipage}{8.5cm}
$^{i}$ X-ray data sources: \\
The count rates CR, and HR1 are from the RASS Bright Source Catalog
(Voges et al. 1997).  All \ax~ are from single power law fits
with \nh~ unconstrained.  For those objects for which no entry was
found, the RX\,J coordinates were taken from accurate optical
positions.

(a) derived from HR1 and HR2 (see Sect. \ref{obs,x}) \\
(b) Boller et al. (1996)\\
(c) RASS observation (Grupe et al. 1997) \\
(d) ROSAT pointed observation retrieved from the archive 

\end{minipage}
\hspace{1.5cm}
\begin{minipage}{6.6cm}
$^{ii}$ Optical data sources: \\
(a) Goodrich (1989) \\
(b) Osterbrock \& Shuder (1982) \\
(c) Rafanelli (1985) \\
(d) McD2.1 March 1997 (see Appendix A) \\
(e) Osterbrock \& Pogge (1985) \\
(f) McD2.1 March 1994 (Grupe et al. 1997, in preparation) \\
(g) Lipari et al. (1993)
\end{minipage}
\end{table*}

The data for the optically selected NLSy1 sample
are presented in Table \ref{res_goodrich} where we list the 
object's common and RX\,J name, percentage polarization and angle, X-ray count
rates and hardness ratio, \ax, \dnh, FWHM(\hb) and [O\,III]/\hb~ ratio.
The polarization data are mean values over the whole observed wavelength range
(Goodrich 1989).  Because soft-X-ray absorption is more probable in this
non-soft-X-ray-selected sample we give the \ax\ determined from an X-ray spectral
fit with $N_{\rm H}$\ unconstrained.  Sources of data are given below the table.

As expected, the optical selection criteria for this sample result in narrower
\hb\ and a wider range of HR1 compared with the X-ray-selected sample.  Only
half of the optically selected sample have steep \ax\, $\gtrsim 1.5$. 

Out of 17 objects, several are polarized at a low level, but
three show continuum p$>$1\%, even $>$2\%: Mkn\,1239, Mkn\,766, and 
IRAS\,15091$-$2107.  As for the highly polarized 
sources of the soft-X-ray sample, these three have FHWM(\hb) typical of the 
unpolarized or low-polarization AGN, but they are among
the most reddened and have among the highest [O\,III]/\hb\ ratios 
(Figs. \ref{hahb_aopt} \& \ref{ewfe2_o3}).  Mkn\,1239 and IRAS\,15091$-$2107
have the largest HR1 of the sample, apparently the result of soft-X-ray
absorption.  Mkn\,783, while not highly polarized, has a high HR1 and soft X-ray
absorption, and the largest [O\,III]/\hb.

\subsection{\label{res,pol}The Polarized AGN}

\begin{table*}
\caption{\label{res_highpol} Properties of the highly polarized AGN.  Those of our
soft-X-ray sample, together with the median and the 90\% ranges
for the whole soft X-ray sample are to the left. The numbers involved 
are given in brackets after the median value. The right side of
the table gives the values for Goodrich's highly polarized NLSy1s.  The following
properties are tabulated:  
optical monochromatic luminosity log\,$\nu L_{V}$\ in Watts, assuming 
$ H_0~=~75~\rm km~s^{-1}~Mpc^{-1}$\ and $q_0 = 0.5$; redshift z; p\%;
FWHM(\hb) in \km, corrected for instrumental resolution;
the intensity ratio [O\,III]/\hb; Fe\,II/\hb;
rest frame equivalent width EW(Fe\,II) in \AA; the total line flux H$\alpha$/H$\beta$\ ratio;
\aopt\ between 4400\AA\ and 7000\AA;
\ax, the X-ray spectral index with $N_{\rm H}$\ unconstrained;
\dnh, the intrinsic soft-X-ray absorption column
density in units of $\rm 10^{20}cm^{-2}$; $\Delta N_{\rm H,opt}$\ is the 
intrinsic H\,I column density calculated from
$E_{\rm B-V}$ (units as for \dnh).
} 
\begin{tabular}{lcccccc|cccc}
\hline
\noalign{\smallskip}
Property & CBS\,126 & IRAS\,1239 & IRAS\,1334 & median & 90\% range 
& & & Mkn\,766 & Mkn\,1239 & IRAS\,1509 \\
\noalign{\smallskip}\hline\noalign{\smallskip}
${\rm log} \nu L_{\rm V}$  & 37.1 & 36.7 & 37.7 & 37.1 (95) & 36.4 - 38.2  
& & & 36.6 & 36.6 &  37.0 \\
z & 0.079 & 0.044 & 0.107 & 0.107 (95) & 0.04 - 0.34 & & & 0.013 & 0.020 &
0.044 \\
\% Pol & 1.26\pl0.13 & 3.77\pl0.20 & 4.63\pl0.08 & 0.34 (43) & $<$ 1\%  
& & & 2.34\pl0.02 & 3.35\pl0.02 & 4.61\pl0.03 \\
FWHM \hb  & 2850\pl200 & 1900\pl150 & 2200\pl200 & 2250 (91) & 1300 -
4200  
& & &  1360\pl150 & 1050\pl150 & 2250\pl200 \\
{[O\,III]}/\hb & 0.29\pl0.04 & 2.00\pl0.14 & 0.13\pl0.02 & 0.31 (87) &
0.05 -
1.40 & & & 4.38\pl0.20 & 3.06\pl0.06 & 1.38\pl0.06 \\
FeII/\hb & 1.2\pl0.1 & 5.5\pl0.3 & 6.5\pl0.1 & 4.0 (87) & 1.0 - 8.0 
& & & 4.5\pl0.1 & 6.3\pl1.1 & 4.4\pl0.1 \\
EW FeII  & 120\pl10 & 175\pl25 & 465\pl15 & 195 (89) & 100 - 360 
& & & 60\pl2 & 210\pl10 & 160\pl10 \\
H$\alpha$/H$\beta$ & 3.7\pl0.3 & 5.9\pl0.4 & 6.2\pl0.3 
& 3.7 (70) & 2.9 - 4.6 & & & 
5.9\pl0.3 & 6.4\pl0.4 & 9.2\pl0.5 \\
$\alpha_{\rm opt}$ & 1.0\pl0.1 & 2.3\pl0.1 & 1.8\pl0.2 & 1.0 (88) & 0.1
- 1.8
& & &  1.9\pl0.1 & 2.5\pl0.1 & 3.6\pl0.1 \\
$\alpha_{\rm X}$ & 1.6$^1$\pl0.1 & 1.7$^2$\pl0.1 & 1.8$^2$\pl0.1 
& 1.9 (95) & 1.5 - 3.2 & & & 
1.7$^3$\pl0.1 & 3.1$^2$\pl0.3 & 1.6$^2$\pl1.0 \\
\dnh & --0.1$^1$\pl0.6 & +1.2$^2$\pl0.3 & $-$0.0$^2$\pl0.0 
& --0.1 (95) & --1.2 - +1.0 & & & 
1.6$^3$\pl0.4 & 4.6$^2$\pl1.8 & 25.8$^2$\pl4.6 \\
$\Delta N_{\rm H,opt}$ & 3 & 21 & 17 & & & & &  17 & 24 & 39 \\
\noalign{\smallskip}\hline\noalign{\smallskip} \\ 
\end{tabular}
\begin{minipage}{8cm}
{\small
X-ray data sources: \\
$^1$ RASS, Grupe et al. (1997). \ax\ assumes $N_{\rm H}$ = $N_{\rm H, gal}$
} 
\end{minipage}
\hspace{1.5cm}
\begin{minipage}{8cm}
{\small
$^2$ Pointed observation. 
Data retrieved from the MPE ROSAT archive \\
$^3$ Very High State from Molendi \& Maccacaro (1994)
}
\end{minipage}
\end{table*}

Two of the three highly polarized AGN of our sample show high and 
wavelength-dependent polarization -- IRAS\,F12397+3333 (Was\,61, 
Wasilewski 1983)
and IRAS\,13349+2438 (Fig. \ref{i1239_pol}).  The IRAS\,13349+2438
polarization has been the subject of intensive investigation as a result
of its membership in a warm IRAS sample (Sect. 1).  While it provided
motivation for the present survey, its inclusion in our statistical
discussion is justified because it was discovered by completely
independent techniques and was not appreciated to be a member of our sample
until much later.  But IRAS\,F12397+3333 is new.  The strong increase
of polarization towards shorter wavelengths is a clear indication of the
dilution of a scattered light AGN spectrum by a redder unpolarized
component -- either reddened AGN light or starlight.  The other highly polarized
source, CBS\,126, appears to be in a different category.  The polarization is 
small and not strongly wavelength dependent if at all.  Unlike the other 
polarized AGN, the broad lines and continuum are not significantly reddened,
and the larger FWHM(\hb) is more typical of normal Seyfert 1
nuclei.  Possibly the polarization arises from transmission of nuclear light 
through well-aligned grains.  For all these reasons, we exclude this source
from detailed discussion.  
The three optically selected NLSy1s with p$>$2\% also show p\% increasing to
short wavelengths, which, in these AGN, arises from dust-scattered continuum
and broad lines (Goodrich 1989).

\begin{figure}
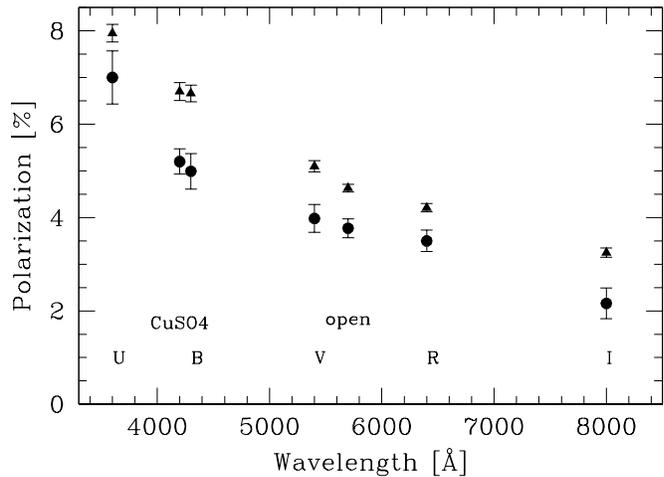

\parbox[t]{8.7cm}{
\clipfig{i1239_pol}{88}{20}{60}{195}{190}
}
\caption[ ]{\label{i1239_pol}
Wavelength dependence of polarization in 
IRAS F12397+3333 (circles) and 
IRAS 13349+2438~ (triangles).  Wavelengths
(\AA) are in the observed frame.
}
\end{figure}

Further details of the X-ray and optical properties of the highly polarized AGN
are given in Table \ref{res_highpol}.  The AGN are approximately in order of
increasing p\% and presumed obscuration.  For comparison, in the same table, we
include the medians and 90 percentile ranges of all measured quantities for
the complete soft-X-ray-selected sample.  We give \ax\ and \dnh\ based upon 
the best available X-ray data.  For all but CBS\,126 and IRAS\,13349+2438, the
X-ray fits with $N_{\rm H}$\ fixed at $N_{\rm H,gal}$\ were unacceptable, so
for these others we have adopted the \ax\ determined from the power-law fits 
with $N_{\rm H}$\ unconstrained.

All scattering-polarized AGN show \aopt\ and \hahb\ indicative of reddening
by dust (Fig. \ref{hahb_aopt}, Table \ref{res_highpol}), and these two
quantities are well correlated.  Except for IRAS\,13349+2438, the degree of
polarization is correlated with reddening, but any such real relation must be 
complicated by differential reddening of scattered and direct light, different
amounts of starlight dilution, and the sensitivity of polarization to details
of the non-spherical, projected geometry.  Greater optical reddening appears to
be well correlated with stronger cold absorption in soft-X-rays as indicated by
\dnh\ (Table \ref{res_highpol}).  IRAS\,15091$-$2107, with nearly 3 magnitudes of
absorption in the optical, is nearly completely absorbed in soft X-rays
below 0.5 keV.
We should be careful here.  The true soft-X-ray
absorption may be even greater than indicated by \dnh\ if a `soft excess' 
is present
at the lowest photon energies -- the kind attributed to a very hot accretion 
disk (Puchnarewicz et al. 1995b; Grupe et al. 1997).

Except for IRAS\,13349+2438, the scattering-polarized AGN all have high
[O\,III]/\hb\ ratios (Fig. \ref{ewfe2_o3}).  Alternatively, one could
describe the highly polarized sources in this figure as having high EW(Fe\,II) 
for a given [O\,III]/\hb\ ratio (\ref {discus,red}).

Like the X-ray-selected AGN as a whole,
the highly polarized AGN show rapid, large-amplitude, time-variability in
soft X-rays.
IRAS\,13349+2438 has shown $\sim$50\% variations over $\sim$6000 s (Brinkmann 
et al. 1996, Brandt et al. 1997).
For IRAS\,F12397+3333 we find $\approx$50\% variations over $\approx$20 ks.
Mkn\,766 varied by a factor 2 in 1000 seconds (Leighly et al.
1996).  Mkn\,1239 varied between the RASS and pointed observations about two
years later.  The count rate doubled and \ax~ changed from 1.69 during the
RASS to 1.94 in the pointed observations, as derived from single power law
fits with \nh\ fixed to \nhgal\ (Rush \& Malkan 1996).
IRAS\,15091$-$2107 has shown a probable variation from 0.2 \cts\ to 0.3 \cts\
over 3 days of ROSAT pointed observations, compared with 0.37 \cts\ in the
RASS 2.5 years earlier.

\section{\label{discus} Discussion}
\subsection{\label{discus,stats}Statistics of Soft-X-ray NLSy1s}


We found 40 of 43 soft-X-ray AGN to have p $<$1\%, 
consistent with no intrinsic polarization.  Most objects 
therefore have a direct, unobscured line-of-sight to the nucleus.
Some arguments in favor of this can be summarized:
\begin{itemize}
\item Lack of cold X-ray absorption along the line-of-sight.  The distribution of
\dnh~ is basically consistent with the errors in \nh\ derived from spectral
fits with  $N_{\rm H}$\ unconstrained (Grupe et al. 1997).
\item Rapid X-ray variability is common in this sample, suggesting an emission 
source less than a few light-days in size (Grupe et al. 1997; see also Boller 
et al. 1993, 1996, 1997). 
This argues for a direct view of a tiny powerful nuclear source of X-rays and
against an extended source such as is expected for a scattering origin.
\item The flat optical spectra and small Balmer line ratios (\hahb) imply no
significant dusty gas.
\item The lack of significant optical linear polarization suggests a
dustless line-of-sight.
\end{itemize}
These arguments for a direct unobscured view support the hypothesis, based
on spectral energy distributions, that we view the inner accretion disk.

Polarization and reddening are usually associated with cold, dusty gas.
What makes our soft-X-ray AGN sample interesting is that these sources have
steep X-ray spectra and show no soft X-ray absorption by cold, neutral matter,
at least in the RASS spectra (Grupe et al. 1997).  
Thus, it is a surprise that we find any polarized
sources at all -- but we find three -- two of which clearly show scattering
polarization and signs of UV-optical obscuration.  These are not the
ones with the narrowest BLR emission lines, or steepest X-ray spectra.
Both IRAS\,F12397+3333 and IRAS\,13349+2438 have FWHM(\hb) close to the
`magical' NLSy1 borderline of 2000 \km.
%

\subsection{\label{discus,pol} The Polarized AGN}


\subsubsection{\label{discus,red}UV-Optical Reddening}

%

Assuming that dust is responsible for the continuum and emission-line reddening
in the highly polarized AGN, where is this dust located?
The two measures of optical absorption -- \aopt\ and \hahb\ -- are
well correlated (Fig. {\ref{hahb_aopt}).  We show on this figure a line 
representing equal emission
line and continuum reddening with respect to typical, essentially unreddened,
values for the X-ray-selected sample as a whole. 
While we have used total H$\alpha$\ and 
H$\beta$\ line strengths to determine \hahb\ in this figure, we note that the 
four most reddened objects have strong narrow line emission.  In de-blending the
H$\alpha$\ complex, we derived ratios 
[N\,II]\,$\lambda\lambda$6548,6584/[O\,III]\,$\lambda$5007 of 0.4 -- 0.6.
While uncertain, comparison of these with values in Koski (1978) and Cohen (1983)
suggests little reddening of this NLR emission.  Less
reddening of NLR H$\alpha$\ and H$\beta$\ would imply 
even larger H$\alpha_{\rm b}$/\hb\ ratios for the broad lines than are shown in the 
figure.  Therefore, while the continuum and emission lines appear about equally
reddened, it seems likely that the BLR is actually more reddened than the 
continuum.  If more extensive investigation shows this to be true, then the dust 
may be more closely associated with the BLR than with a region that obscures
both the BLR and the continuum source.

A consistency check of BLR reddening is to use the line and continuum reddening
determined  from the relation shown in Fig. \ref{hahb_aopt} to correct the
\hb\ intensity, and thus the ratio [O\,III]/\hb.  If the BLR suffers reddening
similar to the continuum, EW(Fe\,II) may not be much affected by reddening.
This correction brings all the highly polarized sources into
line with the low polarization AGN in Fig. \ref{ewfe2_o3}, consistent with the
[O\,III]\,$\lambda$5007 being relatively unreddened, and arising in a more
spatially extended region such as an ionization cone.  
The BLR reddening has another interesting implication.
The highly polarized AGN cover a wide range
of [O\,III]/\hb\ and EW(Fe\,II) and, after applying this 
correction for absorption of \hb\ they follow the  same well known
inverse (`First Principal Component') relation between
[O\,III]/\hb\ and EW(Fe\,II) shown by other AGN (Boroson \& Green 1992; Grupe
et al. 1997, in preparation).  
We note that, for AGN in general, nuclear reddening must contribute to the
scatter in the strong [O\,III]/\hb\ -- EW(Fe\,II) relation and must 
partially obscure it in samples unbiased with respect to reddening.
 
Reddening introduces important biases when QSOs are selected by optical-UV
brightness and color.  It is no accident that three IRAS-discovered QSOs are
among our sample of highly polarized, reddened AGN.  When corrected for reddening
IRAS\,12397+3333 and IRAS\,13349+2438 have log\,$\nu L_{\nu} \sim 38.0$\ (watt).
Similarly log\,$\nu L_{\nu}$\ 
for Mkn\,1239 and IRAS\,15091$-$2107 becomes
37.2 and 38.1.   IRAS\,12397+3333, IRAS\,13349+2438, and IRAS\,15091$-$2107
are then among the most intrinsically luminous AGN at low redshift.

\subsubsection{\label{discus,warm}Warm Absorbers}

Another way to investigate the dusty regions is to compare the 
neutral-hydrogen column densities inferred from the intrinsic optical
absorption ($\Delta N_{\rm H,opt}$) and from the soft-X-ray spectral fitting
(\dnh).  We derived $E_{\rm B-V}$\ from Fig. \ref{hahb_aopt} to calculate
$\Delta N_{\rm H,opt} \sim {\rm k} \times E_{\rm B-V}$\ 
$\rm 10^{20}~cm^{-2}$ (k = 49, Diplas \& Savage 1994).  The uncertainties could
be $\sim 5 \times \rm 10^{20}~cm^{-2}$, judging by the scatter in
Fig. \ref{hahb_aopt}.  These $\Delta N_{\rm H,opt}$\ values are given in Table 
\ref{res_highpol}.
It can be seen that the optical reddening significantly overpredicts
\dnh\ in all cases except perhaps IRAS\,15091$-$2107, where the very high
soft X-ray absorption may be consistent with the optical reddening. 
If we had used  k = 53 (Predehl \& Schmitt 1995; Predehl \& Klose 1996) or
k = 66 (Gorenstein 1975) the discrepancy would have been even greater.
It is also probable that less-reddened, scattered light contributes to the
optical spectra (Sect. \ref{discus,scat}).  In this case the true reddening
along a direct path to the nucleus, and hence $\Delta N_{\rm H,opt}$,
would be even greater.

The relatively low soft X-ray absorption could be because the X-rays arise in
a region more spatially extended than the UV-optical.  However, the large
amplitude and rapid variability (Sect. \ref{res,pol}) provides evidence that
all or most of the X-ray emission arises within regions light-days to 
light-months in size.  This, together with the high brightness in soft X-rays,
argues that we have a direct view to the center
in X-rays rather than a view via a parsec-scale scattering region.
The relatively low soft X-ray absorption could be because the dust-to-gas
ratio is high, or, more likely in this nuclear environment, the dusty gas
is at least partially
ionized and therefore has greater transparency to soft X-ray photons.
The dusty gas is probably related to the `warm absorbers' detected in
the O\,VII and O\,VIII K-shell edges near 0.7 -- 0.8\,keV.
This ionized gas has now been detected in IRAS\,13349+2438 (Sect.
\ref{discus,scat}), and Mkn\,766 (Leighly et al. 1996).
Warm absorbers provide a
natural tie-in for the UV-optical reddening, X-ray variability, and bright,
unabsorbed soft X-ray spectrum.
In fact, for IRAS\,F12397+3333, neutral hydrogen absorption plus a single
power-law does not provide a good fit to the X-ray data (Grupe et al. 1997).
In the residuals we find a depression near 0.8 keV -- an indication of
O\,VII and O\,VIII absorption edges.  The X-ray spectrum and
variability will be discussed in more detail together with the results of our
spectropolarimetry for this AGN (Wills et al. 1997, in preparation).

Recently, Leighly et al. (1997) show that
AGN with warm absorber features in their X-ray spectra tend to be
significantly polarized.
Related to this, Reynolds (1997) also report that Seyfert 1s with UV
absorption also showed X-ray warm absorption.

Walter \& Fink (1993) should be given credit for pointing out that some
reddened, highly polarized NLSy1s showed anomalously low
$\nu$F$_\nu$(1375\AA) on a plot of $\Gamma$\ vs.
$\nu$F$_\nu$(1375\AA)/$\nu$F$_\nu$(2 keV) (their Fig. 8).
The low UV flux (by $\sim$30)
could be explained on
the basis of their observed optical reddening, but they found little or no
soft X-ray absorption.  All except one are NLSy1s, and now known to show warm
X-ray absorption.  They are IRAS\,13349+2438, Mkn\,766, NGC\,4051
(Table \ref{res_pol}, Komossa \& Fink 1997),
MCG-6-30-15 (FWHM(\hb) $\sim$1700 \km, Pineda et al. 1980, Reynolds et al. 1997),
Akn\,564 (Table \ref{res_goodrich}).
The other NLSy1s discussed here -- IRAS\,12397+3333,
IRAS\,15091$-$2107, and Mkn\,1239 -- follow the same trend.  When corrected for
a reddened continuum all these  NLSy1s show the strong increase of $\Gamma$\
with $\nu$F$_\nu$(1375\AA)/$\nu$F$_\nu$(2 keV) that led Walter \& Fink to
argue for a UV to soft X-ray bump in Seyfert 1 nuclei.

\subsubsection{\label{discus,scat}Scattered Light Models}



We have some evidence that the scattering geometry for NLSy1 may be 
axisymmetric, and this is certainly true for many other AGN.
The polarization (E-vector) is parallel to the major axis of the galaxy in
the case of IRAS\,13349+2438, and for Mkn\,766 and Mkn\,1126 the polarization 
is perpendicular to the elongated radio structure thought to define the jet 
direction -- or angular momentum of the central engine (Ulvestad et al. 1995).
Mkn\,766 shows an `ionization cone' in the jet direction (Wilson
1997).  If we can assume axisymmetry we can relate scattered and direct
views to the geometry of the central engine -- for example, accretion disk
geometry and kinematics.

IRAS\,13349+2438 is the best studied of the highly polarized AGN
(Wills et al. 1992b; Brandt et al.  1996, 1997; Brinkmann et al. 1996; 
Hines 1994).
In this AGN the contribution of host-galaxy starlight is $<$15\%, so 
the wavelength dependence of polarization (Fig. \ref{i1239_pol}) is
explained by
dilution of polarization by reddened direct AGN light.  Therefore we see
the AGN from two directions -- a direct view, and a view from the vantage point
of the scatterers.  The polarization E-vector is parallel to the host galaxy's
major axis, suggesting an axisymmetric geometry with obscuring dust in the 
plane of the galaxy, probably a dusty torus.
\hb\ and H$\alpha$ may be slightly narrower in scattered than in
total light (Hines 1994).
X-ray emission in the ROSAT band is strong and rapidly variable, and therefore
not seen reflected from a large (parsec-scale) scattering region, but
is instead nuclear light seen directly.  
The steep X-ray spectrum, lack of neutral absorption in X-rays, but strong
UV-optical reddening, led Brandt et al. (1996, 1997)
to suggest that ionized, dusty gas absorbs the nuclear light, thus leading to
the first clear detection of  warm absorbers in a luminous QSO.
Both the fact that the central AGN is still optically visible, and that the 
absorption is warm,
suggest a line-of-sight to the center that grazes the dusty torus: the 
absorption optical depths are significant but not extremely high; grains are not
likely to have formed in the ionized gas, but pre-existing grains could have
evaporated from the inner torus on exposure to the nuclear continuum.  

A similar scattered-light--dusty torus model
is a plausible explanation of our second highly polarized object,
IRAS\,F12397+3333.  In this case direct sky survey images show that unpolarized
host-galaxy starlight is likely to contribute to the wavelength dependence of
p\% (Fig. \ref{i1239_pol}).  Warm absorbers provide a
natural explanation for the UV-optical reddening, X-ray variability, and bright,
essentially unabsorbed soft X-ray spectrum.


Goodrich's (1989) spectropolarimetry suggests a similar scattering
geometry for the three highly polarized AGN in his sample of 17 optically
selected NLSy1s (Tables \ref{res_highpol} and \ref{res_goodrich}), and our
investigation of their UV-optical reddening and X-ray spectra suggest both cold
and warm absorbers along a direct path to the nucleus.

The prototypical NLSy1, I\,Zw\,1, is an
interesting example.  Because of its extreme NLSy1 properties, NLSy1 are
sometimes called `I\,Zw\,1' AGN.
It was too weak to be included in our X-ray sample because of anomalously
high Galactic HI absorption and variability.  Its HR1 was +0.49 during
the RASS, but $-$0.25 during later pointed observations (Boller et al. 1996).
%
It shows significant scattering polarization,
with polarized continuum and
broad lines  (1.7\%, Smith et al. 1997) variable in position angle and
wavelength-dependence on several-year time scales -- maybe shorter --
and possibly significant optical reddening, E(B-V) $\sim 0.2$.  Its steep
soft-X-ray spectrum (\ax $\sim$ 2.0) is rapidly variable, with very small
intrinsic cold absorption ($1.5 \pm 0.7 \times 10^{20}$ cm$^{-2}$).
If reddened, this is likely to be another example of a warm absorber -- and
on axisymmetric Unified Schemes, would therefore be a candidate for a
torus-grazing line-of-sight -- a potentially powerful illustration that steep
X-ray spectra need not be observed from the same direction as the narrow
\hb\ (and other `I\,Zw\,1' properties that go along with these properties).

\subsubsection{\label{discus,nature}The Nature of NLSy1s}

In principle, comparison of scattered and direct views allows a test for the
predicted axisymmetry in kinematics, UV- and X-ray emission (Sect.\ref{intro}).

The important conclusion from Goodrich's (1989) spectropolarimetry is that the
narrow H$\beta$\ in NLSy1 is more highly polarized than [O\,III], whose polarization
is often unmeasureably small; \hb\ is narrow not because it arises in the NLR;
it is produced in separate, higher density gas.

Assume that polarized \hb\ represents a scattered light, polar view.
Broader polarized \hb\ than that seen in total light
could then be interpreted as evidence for scattering by electrons in hot ionized
nuclear gas (Goodrich 1989), or for a dust-scattered view of a high-velocity inner
BLR and central continuum source that are obscured in a direct
high-inclination, view.  Narrower polarized \hb\ could be interpreted as 
arising from a polar view of a flattened BLR.

In general, in the $\sim$5 cases where a comparison is possible, 
Goodrich does not find significantly broader \hb\ in polarized light, and
therefore favors scattering by cool dusty gas.
This assumes that a predominantly direct view of the center is seen in total
light; without knowing p\% for the scattered light alone, we cannot be sure.
Also from the similar widths in polarized and total light,
we could argue that a higher-velocity, inner BLR is not hidden by
the thicker equatorial regions of a dusty torus, and that there is no 
evidence for an anisotropic velocity field.
Mkn\,1239 is an interesting exception.  Goodrich finds broad, red wings as an
additional feature of the H$\beta$\ and H$\alpha$\ broad lines in polarized
flux, which he interprets as arising in a separate electron-scattering region.

AGN like IRAS\,13349+2438 and IRAS\,12391+3333 of our sample, and three of the
Goodrich sample, show significant scattered light views of an optical
NLSy1 BLR spectrum (p$_{max} \sim 9$, 7, 6, 6, 8\%), but a direct, higher
inclination view of a luminous soft X-ray source with steep 0.2--2 keV spectrum.
This rules out the hypothesis in
which the narrow \hb\ and steep \ax\ belong exclusively to a pole-on view of an
optically thick accretion disk with coplanar BLR (e.g., Boller et al. 1996).
Mkn\,1239 is especially extreme, with \ax $\sim$3 and FWHM(\hb) in polarized
and total light of $\sim$1500 \km\ and $\sim$1000 \km.



\subsection{\label{discus,us}Unified Schemes}


The similarity of the strong, variable, steep X-ray spectra, and narrow \hb\
in both highly polarized
and low polarization AGN support a Unified Scheme in which the low radial BLR velocity
dispersion and steep X-ray spectra are intrinsic properties that are not highly
anisotropic; the differences are all attributable to orientation-dependent 
scattering and obscuration.  When corrected for optical reddening the highly polarized
objects fit the inverse correlation between Fe\,II and [O\,III] strengths seen
for the unreddened AGN, and even the optically-selected NLSy1s show steep X-ray
spectra after correction for cold absorption.  
The highly polarized AGN lead to a picture of partial obscuration of the central
continuum and BLR, with polarized light reaching us from scatterers that have a
less-obscured view of the center.
While the case for an axisymmetric Unified Scheme is based on polarization
alignment in only a few AGN, the partial obscuration of a direct view of the AGN
by dusty, ionized gas suggests a line-of-sight grazing the dusty torus.
All the above leads us to consider a unified picture in which all NLSy1s have
dusty tori and scatterers, with some lines-of-sight unobscured.

Our sample of 43 comprises all AGN in a ROSAT X-ray source sample, complete to a
given X-ray count rate and hardness ratio.  All the AGN
(except IC\,3599) have prominent AGN-like optical continua and broad
H$\beta$.  There are no classical Seyfert 2s in the sample at all,
demonstrating that our view towards their nuclei is opaque in soft X-rays.  Our
sample includes only two out of 43 AGN that are clearly scattering-polarized,
demonstrating that warm, dusty absorbers are rarely found  along sight-lines
without accompanying cold absorbing gas.  In fact, the soft-X-ray
absorption seen in the highly polarized AGN of the X-ray and optically selected samples,
while small,
does correlate well with optical reddening, but we suggest that an additional
\nh\ of 1.5 -- 2 $\times 10^{21}$\, cm$^{-2}$\ is accounted for by warm
X-ray absorbers.

%

Assuming axisymmetry, the low polarization AGN represent unobscured views from within an
ionization-scattering cone.  Several sources in both samples show observed white
light p $<$1\%.  Goodrich (1989) found that some lower polarization
objects show p\% increasing to shorter wavelengths with significant continuum
and broad-line polarization, indicating that a number of sources in both
samples show similar scattering geometry.  Their smaller polarizations may arise
from a smaller scattering angles, and greater dilution by unpolarized AGN and
host-galaxy light.
At higher inclination angles,
partially ionized dusty gas absorbs the UV, soft X-rays and $\sim$0.7 keV
X-rays.  It doesn't take much dusty gas, like that of our interstellar medium,
to completely block the UV and soft-X-rays,
so it is probably not surprising that only 5\% of our sample are highly polarized
warm absorbers.  As the inclination increases the thicker dusty torus blocks
all but the harder X-rays (5-15 keV, e.g. Comastri et al. 1995).  At higher
inclinations, IRAS\,12397+3333, Mkn\,766, Mkn\,1239, and IRAS\,1509$-$2107
would appear as Seyfert 2s (like NGC\,1068, discussed below).
Strong-Fe\,II, weak-[O\,III] AGN like
IRAS\,13349+2438 would appear as galaxies with weak emission lines,
detectable in hard X-rays and warm mid-infrared surveys.
Deep X-ray spectral surveys of complete infrared-selected
AGN are needed to understand the covering by cold and warm X-ray absorbers.

The archetype buried Seyfert 1 nucleus, revealed in the polarized light spectrum
of the classical Seyfert 2,
NGC\,1068, may simply be a more edge-on view of a NLSy1, with
the UV-optical and soft X-rays seen only in scattered light.  In this case,
a spatially resolved dust-scatterer sees a low inclination view with
FWHM(\hb) $\sim$ 2900 \km, but electron scatterers within a few parsecs of the
nucleus produce FWHM(\hb) $\sim$ 4480 \km\ presumably the result of thermal
broadening at $\sim$ 10$^6$ K (Miller et al. 1991).
Electron-scattering is wavelength independent, so
an intrinsic, polar view is seen in soft X-rays, with \ax\ $\sim 2$\ (Marshall et
al. 1993; Pier et al. 1994).  Unlike the result for most NLSy1s with purported
low-inclination views,
the X-ray spectrum shows no time-variability over several years
(Smith et al. 1993), consistent with scattering of X-rays from
a parsec-scale region.

\subsection{\label{discus,ic}IC 3599} 

IC\,3599 is the only AGN of the soft X-ray selected sample to be
classified as a Seyfert 2 galaxy (Grupe et al. 1995).  Unified Schemes
suggest that scattered light polarization might be detectable.
We find no significant polarization, but some optical absorption
may be present (Fig. \ref{hahb_aopt}), a result that may favor the
starburst classification suggested by Bade (1993).  This object would not 
usually fit the criteria for inclusion in the soft X-ray 
sample; it just happened to be in an X-ray outburst at the time
of the RASS (Brandt et al. 1995; Grupe et al. 1995).

\section{\label{sum} Summary}

We have surveyed the optical linear polarization of the 43 northern objects of
our completely-identified sample of bright soft-X-ray-selected ROSAT AGN.
All except one are Seyfert 1 nuclei, with median FWHM(\hb) of 2250\,\km.
Most (40) are of low polarization with p $<$0.5\% -- 1\%, and have
flat blue continua and small Balmer decrements.  This supports the suggestion
from rapid X-ray variability, disk-like spectral energy distribution, and 
lack of cold X-ray absorption, that we are viewing a bare AGN disk (Grupe et al.
1997). 

Three AGN show significant polarization, with no preference for the narrowest \hb\
or steepest \ax.  One, CBS\,126, has p $\sim$1.3\% with no clear wavelength
dependence.  Two, IRAS\,13349+2438 and IRAS\,12397+3333, show high degrees of
polarization increasing to short wavelengths, and significant optical reddening.
In IRAS\,13349+2438 a dusty torus dims the central luminous AGN, revealing a 
less-reddened polar view of the continuum and BLR in scattered light (Wills et al.
1992b).  This partial dust covering forms the basis for Unified Schemes -- accounting
for very different obscured or unobscured views of the same central engine.
Brandt et al. (1996) explained its rapidly-variable and bright X-ray spectrum, 
lack of soft-X-ray absorption, but reddened UV-optical spectrum,
by dusty, ionized gas (warm absorbers) along the line-of-sight to the
center.  The optical polarization and reddening, together with its rapidly-variable
and steep X-ray spectrum, suggest the same scattering--warm-absorber picture for
IRAS\,12397+3333.

We have compared the polarization, spectroscopic and X-ray properties of our two
highly-polarized sources with three discovered by Goodrich (1989) among a 
heterogeneous sample of 17 optically-selected NLSy1s.
Only about half the optically-selected sample have steep \ax, $\gtrsim 1.5$.
There appear to be correlations among
indicators of dust and cold X-ray absorption: p\%, \aopt, \hahb, and \dnh.
  When
the \hb\ strength is corrected for reddening, the highly polarized objects have similar
[O\,III]/\hb\ to the low polarization AGN, and show the same `First Principal
Component' [O\,III]/\hb\ -- EW(Fe\,II) anticorrelation.
However the neutral soft X-ray absorption is overpredicted by the optical
reddening by $\sim$5 $\times 10^{20}$\,cm$^{-2}$.  So, in line with the warm
absorbers seen in IRAS\,1334+2438 and Mkn\,766, dusty partially ionized gas is
almost certainly absorbing nuclear light.

The intrinsic properties of the high and low polarization AGN are similar,
including narrow \hb\ and steep \ax; the differences are all attributable
to orientation-dependent scattering and obscuration, and this is supported by
the periscopic geometry deduced for individual polarized AGN.  Thus an orientation
Unified Scheme may link all the AGN of our sample, and NLSy1s in general. 

\acknowledgements{We gratefully acknowledge Mike Brotherton, Karen Leighly, and 
the referee for valuable comments that improved the paper.
We thank Darrin Crook, Mike Ward, David Doss, Ed Dutchover,
Marian Frueh, and Jerry Martin
of McDonald Observatory for instrumental and observing help.  We also
want to thank Norbert Bade (Sternwarte Hamburg), Wolfram Kollatschny 
(Universit\"ats-Sternwarte
G\"ottingen), and Liz Puchnarewicz 
(University College London), for supplying their spectra of CSO\,150, NGC\,4593, and 
RE\,J1034+39; also Laura Kay, Sally Stephens, and Hien Tran, for finding a
spectrum for MS\,0919.3+5133.
We also thank 
Neil Brandt for sending his preprints on IRAS\,13349+2438, and Alice Breeveld for her
prepint on RX E1034+39.
DG and BJW were supported by a grant from the Space Telescope Science Institute
(GO-06766) and NASA Long Term Space Astrophysics grant NAG5-3431.
This research has made use of the NASA/IPAC Extragalactic Database (NED)   
   which is operated by the Jet Propulsion Laboratory, California Institute   
   of Technology, under contract with the National Aeronautics and Space      
   Administration.                                                            
    
}


\newpage

\def \charthoffset  {\hspace{0.2cm}}
\def \charthsep     {\hspace{0.3cm}}
\def \chartvsep     {\vspace{0.3cm}}
\newcommand{\putchart}[1]{\clipfig{#1}{58}{10}{10}{280}{195}}
\newcommand{\chartline}[3]{\parbox[t]{18cm}{
 
\noindent\charthoffset\putchart{#1}\charthsep\putchart{#2}\chartvsep\putchart{#3}\chartvsep}}

%
%
%

.
\eject

\newpage

\begin{minipage}[t]{18cm}
\appendix
\section{Optical Spectra}
We display the optical spectra of the highly polarized sources of the ROSAT soft-X-ray 
sample, CBS\,126, IRAS\,F12397+3333, and IRAS\,13349+2438, and of Goodrich's sample,
Mkn\,766, Mkn\,1239, and IRAS\,15091$-$2107, in order of increasing p\% and Balmer
decrement within each sample, and approximately in order of increasing obscuration as a
whole.  The left panel shows the
whole wavelength range covering H$\beta$\ and H$\alpha$.  The middle panel illustrates
the Fe\,II blend contribution
in the \hb\ region:  the observed spectrum, (displaced in $F_{\lambda}$), the 
Fe\,II-subtracted spectrum (not displaced), and the Fe\,II template with its zero level
shown as a dotted line.  The right panel shows the H$\beta$\ and 
[O\,III]\,$\lambda$4959,5007 profiles, illustrating the removal of NLR emission from
H$\beta$. 
All spectra were obtained at the McDonald Observatory at the 2.1m Struve telescope
in March 1997, except the spectrum of IRAS\,13349+2438, which was obtained at the 2.7m
Smith telescope in March 1987 (Wills et al. 1992b). 
The spectra are plotted as flux density $F_{\lambda}$\ in units of
$\rm 10^{-19}~W~m^{-2}~\AA^{-1}$\ vs. $\lambda$\ in \AA.    The absolute flux density 
calibration is uncertain.
\end{minipage}

\vspace*{1.5cm}

\chartline{cbs126_res}{cbs126_fe2}{sub_cbs126}

\chartline{ir1239_res}{ir1239_fe2}{sub_ir1239}

\chartline{ir1334_res}{ir1334_fe2}{sub_ir1334}

\begin{figure*}
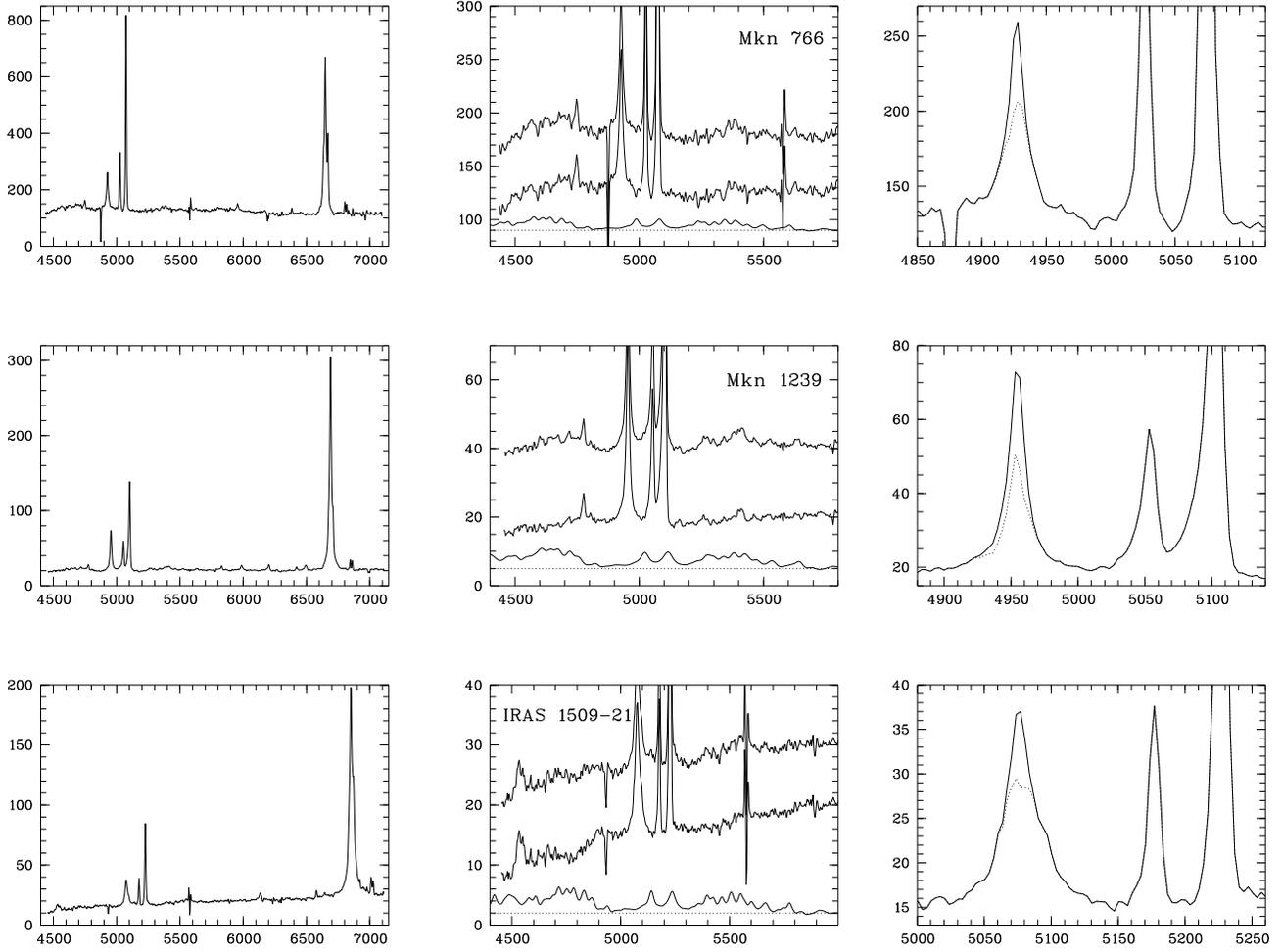

\chartline{mkn766_res}{mkn766_fe2}{sub_mkn766}

\chartline{mkn1239_res}{mkn1239_fe2}{sub_mkn1239}

\chartline{ir1509_res}{ir1509_fe2}{sub_ir1509}

\caption[ ]{Optical spectra of the highly polarized AGN}
\end{figure*}
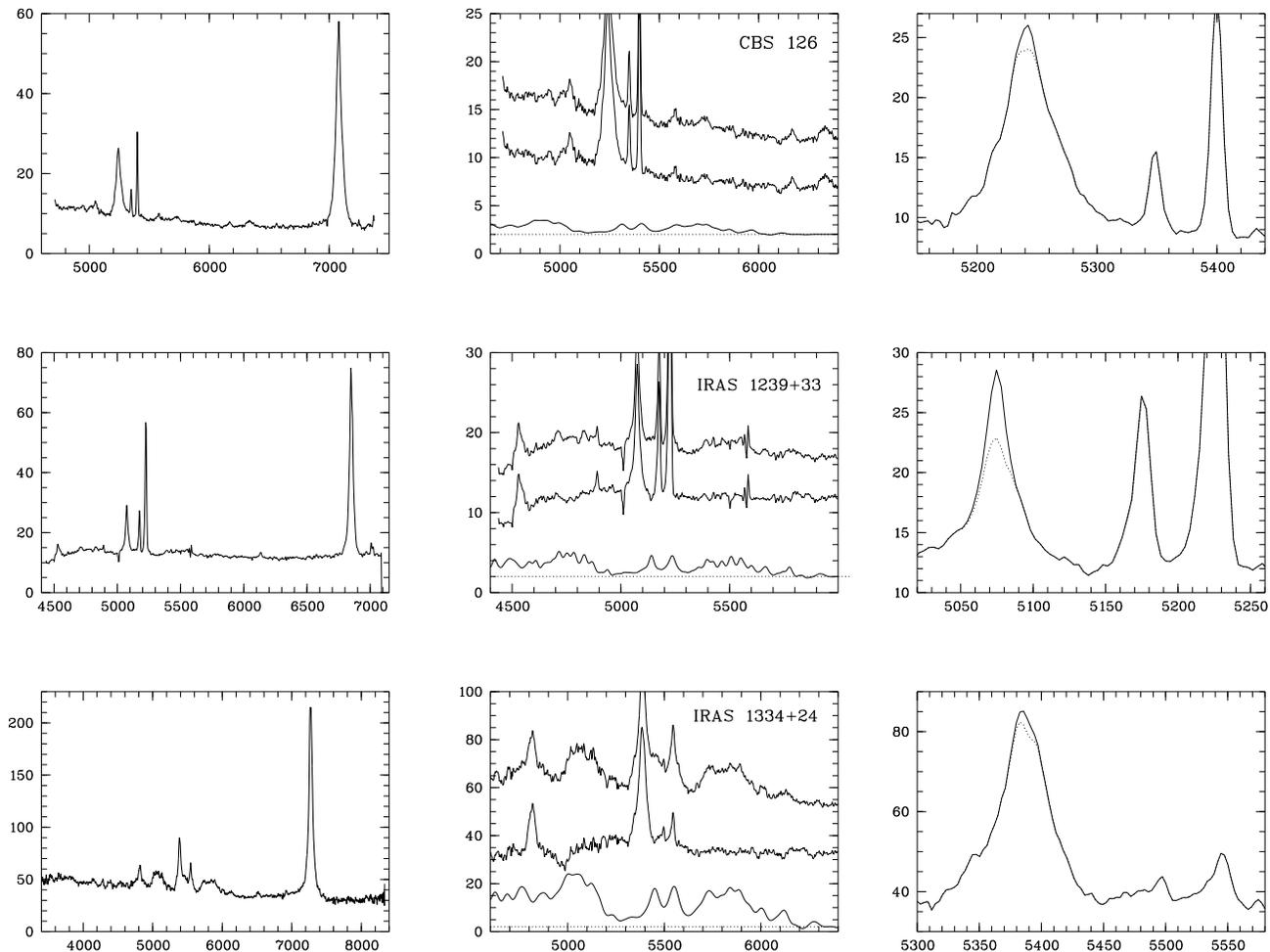
\end{document}